\documentclass[pdflatex,sn-mathphys-num]{sn-jnl}

\usepackage{graphicx}%
\usepackage{multirow}%
\usepackage{amsmath,amssymb,amsfonts}%
\usepackage{amsthm}%
\usepackage{mathrsfs}%
\usepackage[title]{appendix}%
\usepackage{xcolor}%
\usepackage{textcomp}%
\usepackage{manyfoot}%
\usepackage{booktabs}%
\usepackage{algorithm}%
\usepackage{algorithmicx}%
\usepackage{algpseudocode}%
\usepackage{listings}%

\usepackage{url} 
\usepackage{epsfig}
\usepackage{subcaption}
\usepackage{calc}
\usepackage{multicol}
\usepackage{pslatex}
\usepackage{apalike}

\usepackage{verbatim}
\usepackage[colorinlistoftodos]{todonotes}

\newcommand{\footurl}[1]{\footnote{\url{#1}}}

\newcommand{\furl}[2]{\footnote{#1: \url{#2}}}

\newenvironment{smallverbatim}
  {\verbatim\normalfont\fontsize{8}{8}\ttfamily\selectfont}
  {\endverbatim}

\theoremstyle{thmstyleone}%
\theoremstyle{thmstyletwo}%

\theoremstyle{thmstylethree}%

\begin{document}

\title[Article Title]{Semantic Web And Software Agents -- A Forgotten Wave of Artificial Intelligence?}


\author*[1]{\fnm{Tapio} \sur{Pitkäranta}}\email{tapio.pitkaranta@iki.fi}

\author[1,2]{\fnm{Eero} \sur{Hyvönen}}\email{eero.hyvonen@aalto.fi}


\affil*[1]{\orgdiv{Department of Computer Science}, \orgname{Aalto University}, \orgaddress{\street{Konemiehentie 2}, \city{Espoo}, \postcode{02150}, 
\country{Finland}}}

\affil[2]{\orgdiv{Helsinki Centre for Digital Humanities (HELDIG)}, \orgname{University of Helsinki}, 
\country{Finland}}







\abstract{The history of Artificial Intelligence (AI) is a narrative of waves—rising optimism followed by crashing disappointments. AI winters, such as the early 2000s, are often remembered as barren periods of innovation. This paper argues that such a perspective overlooks a crucial wave of AI that seems to be forgotten: the rise of the Semantic Web, which is based on knowledge representation, logic, and reasoning, and its interplay with intelligent Software Agents. Fast forward to today, and ChatGPT has reignited AI enthusiasm, built on deep learning and advanced neural models. However, before Large Language Models (LLMs) dominated the conversation, another ambitious vision emerged—one where AI-driven Software Agents autonomously served Web users based on a structured, machine-interpretable Web. The Semantic Web aimed to transform the World Wide Web into an ecosystem where AI could reason, understand, and act. Between 2000 and 2010, this vision sparked a significant research boom, only to fade into obscurity as AI’s mainstream narrative shifted elsewhere. Today, as LLMs edge toward autonomous execution, we revisit this overlooked wave. By analyzing its academic impact through bibliometric data, we highlight the Semantic Web’s role in AI history and its untapped potential for modern Software Agent development. Recognizing this forgotten chapter not only deepens our understanding of AI’s cyclical evolution but also offers key insights for integrating emerging technologies.}




\maketitle


\section{Introduction}
\label{sec:introduction}

The notion of Artificial Intelligence (AI) as a distinct discipline within Computer Science was introduced in 1956 at a workshop at Dartmouth College by early pioneers such as John McCarthy, Marvin Minsky, and others. Since then, the history of AI has experienced recurring cycles of enthusiasm and subsequent disillusionment, commonly referred to as ``AI summers'' and ``AI winters'' \cite{McCorduck-2004}. An AI winter refers to a period of reduced funding, interest, and progress in AI research, typically following unmet expectations or technological limitations. In contrast, an AI summer denotes a phase of heightened enthusiasm, investment, and innovation, often driven by breakthroughs in algorithms, computing power, or applications. These cycles are associated with two main approaches to creating AI: symbolic (``white box'') approaches, based on explicit knowledge representation, logic, reasoning, and search, and sub-symbolic (``black box'') approaches, based on neural networks, machine learning, and statistical models.

Following the first AI summer in the 1960s, the first AI winter occurred in the 1970s after it was argued in \cite{minsky-papert-1969} that sub-symbolic neural networks (then based on single-layer perceptron) would never be useful for solving real-world tasks. As a response, the symbolic approach, including knowledge-based expert systems and logic programming, fuelled the next AI summer in the 1980s. However, challenges in, e.g., knowledge representation and acquisition \cite{mettrey-1987} soon  led to another AI winter. Research on multi-layer sub-symbolic neural models and learning systems continued despite—or perhaps because of—the shortcomings of the knowledge-based approach. By expanding machine learning models through deep learning, employing new training algorithms such as backpropagation, introducing novel neural architectures and non-linear activation functions, leveraging Big Data from the Web and available databases for training, , and by using new efficient AI chips for parallel neural computations AI has entered an unprecedented summer in the 2020s \cite{kelleher_deep_2019}. A prominent research topic today, aimed at addressing challenges in neural systems such as hallucinations, is the combination of symbolic and sub-symbolic approaches in neuro-symbolic (hybrid) AI systems \cite{Hitzler-sarker-2022,Bhuyan-et-al-2024}.

The early 2000s are often labelled as an AI winter, with little recognition of the relevant AI research conducted during that time. However, this narrative overlooks significant developments from that era, including the emergence of the Semantic Web, which gained momentum in 2001 \cite{berners_lee_2001,BernersLee2003SemanticWebWave}. Although rooted in knowledge representation, logic, reasoning, and ontologies—core subjects of traditional symbolic AI—this branch of research is typically not classified as part of AI. For example, in the authoritative ACM Computing Classification System\furl{ACM Computing Classification System}{https://dl.acm.org/ccs} of Computer Science, AI is a major first-level category under Computing Methodologies.

\begin{smallverbatim}

Computing Methodologies
   >Artificial Intelligence, 
\end{smallverbatim}

while Semantic Web is only mentioned under the minor category under Information Systems:

\begin{smallverbatim}

Information Systems
   >World Wide Web
      >Web data description languages
         >Semantic web description languages   
         
\end{smallverbatim}

\begin{figure}[!h]
    \centering
\includegraphics[width=1.0\textwidth]{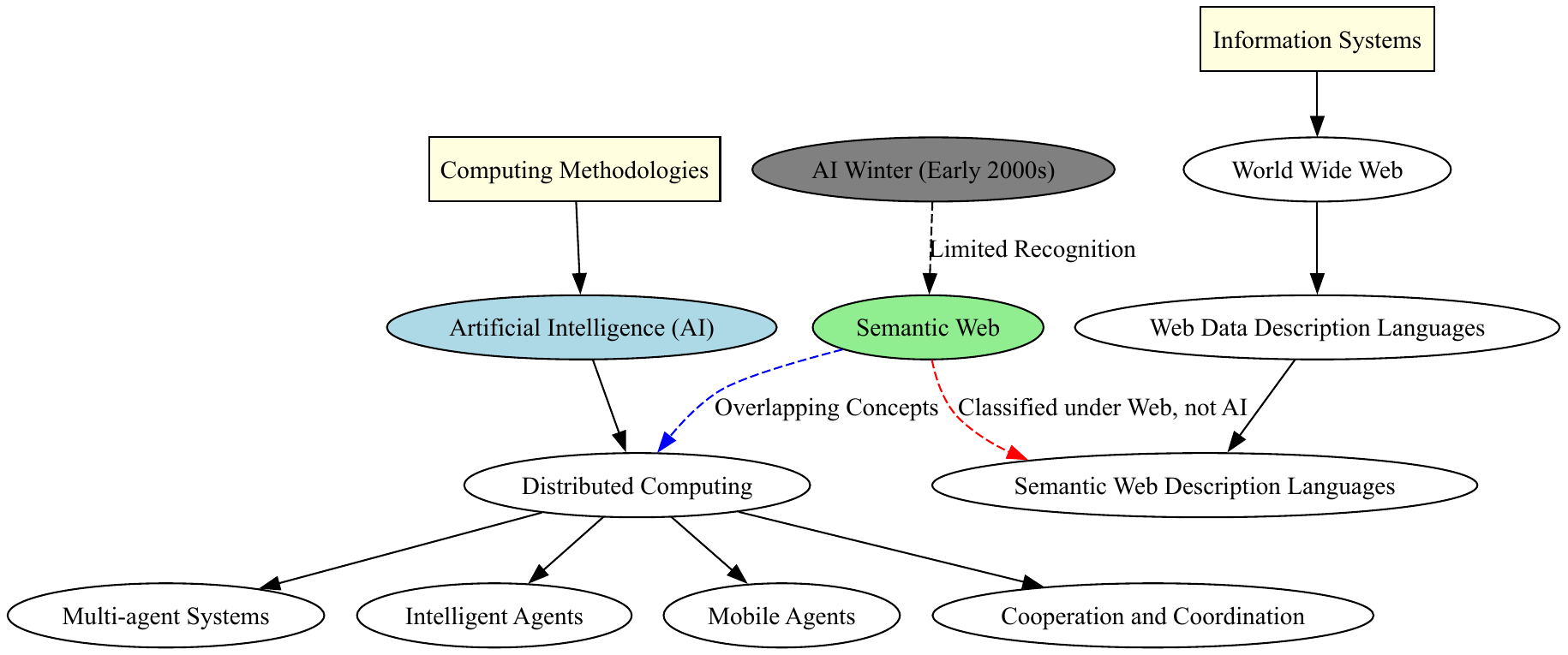}
	\caption{Categorization issue: Semantic Web research vision had strong artificial intelligence focus but it is categorized under WWW branch}
\label{fig:categorizaion-issue}
\end{figure}
According to the original Semantic Web vision outlined in 2001 \cite{berners_lee_2001}, the Semantic Web is centered around intelligent agents on the Web. The category of Intelligent Agents can be found as a minor subcategory under Artificial Intelligence. This categorization issue is also illustrated in Figure~\ref{fig:categorizaion-issue}.

\begin{smallverbatim}

Computing Methodologies
   >Artificial Intelligence, 
      > Distributed computing
         > Multi-agent systems
         > Intelligent agents
         > Mobile agents
         ...
\end{smallverbatim}

Even review papers on the Semantic Web do not focus on the AI or Software Agent research conducted during that period within the Semantic Web domain. For example, Hitzler \cite{hitzler2021review} provides a comprehensive review of the Semantic Web, detailing its evolution, core principles, and ongoing challenges. The paper highlights how the Semantic Web extends the traditional Web by incorporating structured and linked data to enhance machine interoperability. Key technologies such as RDF (Resource Description Framework), OWL (Web Ontology Language), and SPARQL are discussed, emphasizing their roles in enabling semantic interoperability. In addition, the paper explores real-world applications in domains such as AI, knowledge graphs, and data integration. Hitzler also addresses challenges such as scalability, reasoning complexity, and data quality, underscoring the need for continued research and development to fully realize the Semantic Web's potential.

The rapid advancement of Generative Artificial Intelligence (GAI) and the widespread adoption of Large Language Models (LLMs) were catalyzed by the introduction of the Transformer architecture, which enabled more efficient training and superior contextual understanding in deep learning models \cite{vaswani2017attention,kelleher_deep_2019}. This breakthrough laid the foundation for applications such as ChatGPT, which demonstrated the potential of LLMs in generating human-like text, leading to a surge in research and commercial interest. More recently, the focus has expanded beyond standalone LLMs to the integration of Software Agents and agentic capabilities, where models exhibit autonomy, goal-directed behaviour, and enhanced reasoning abilities \cite{reed2022generalist,yao2022react}.
These agentic systems, which incorporate planning, memory, and interaction mechanisms, are now being extensively studied to augment the functionality of LLMs, enabling them to operate dynamically in complex environments, solve multi-step problems, and interact more effectively with users and digital ecosystems.

It appears that for some reason the Semantic Web is a forgotten wave of AI although the current AI trend is very much happening on the Web. To study and analyze this argument, we review previous trend analyses related to the Semantic Web and AI and present statistical analyses of academic literature on these topics.


The remainder of this paper is structured as follows.
\begin{itemize}
    \item Section~\ref{sec:ai-history} provides an overview of the AI hype cycle, detailing the recurring patterns of ``AI summers'' and ``AI winters''. We review and synthesize insights from previous studies on AI history and introduce a normalized ``temperature'' scale to quantify these fluctuations.
    \item Section~\ref{sec:semantic-web-agents} explores the concept of the Semantic Web, its foundational vision, and its relationship with AI-driven Software Agents.
    \item Section~\ref{sec:academic-statistics} presents an analysis of bibliometric data and academic trends related to Semantic Web and Software Agent research. We examine search term frequencies and notable researcher contributions to illustrate the presence of the ``Forgotten AI Wave''.
    \item Section~\ref{sec:discussion} provides a critical discussion of our findings, highlighting the broader implications of the legacy of the Semantic Web, its challenges, and its potential resurgence in modern AI research.
    \item Section~\ref{sec:conclusions} summarizes the key takeaways from our study and outlines directions for future research at the intersection of the Semantic Web and contemporary AI advancements.
\end{itemize}


\section{AI Summers and Winters} \label{sec:ai-history}

In this section, we first discuss the general hype cycle model in Section~\ref{sec:hype-cycle}, a framework used to describe the waves of enthusiasm followed by subsequent disappointment that technological fields often experience. This pattern, commonly referred to as the ``hype cycle,'', has been observed in various industries and technologies.

We then review and synthesize illustrations of AI history in Section~\ref{sec:ai-history-synthesis}, exploring the alternate periods of ``AI winters'' and ``AI summers'' and placing them on a normalized scale. These phases highlight the cycles of overenthusiasm, disillusionment, and eventual productivity, from the early promises of symbolic AI to the breakthroughs of the deep learning revolution.

\subsection{The Hype Cycle} \label{sec:hype-cycle}

\begin{figure}[!h]
    \centering
\includegraphics[width=0.9\textwidth]{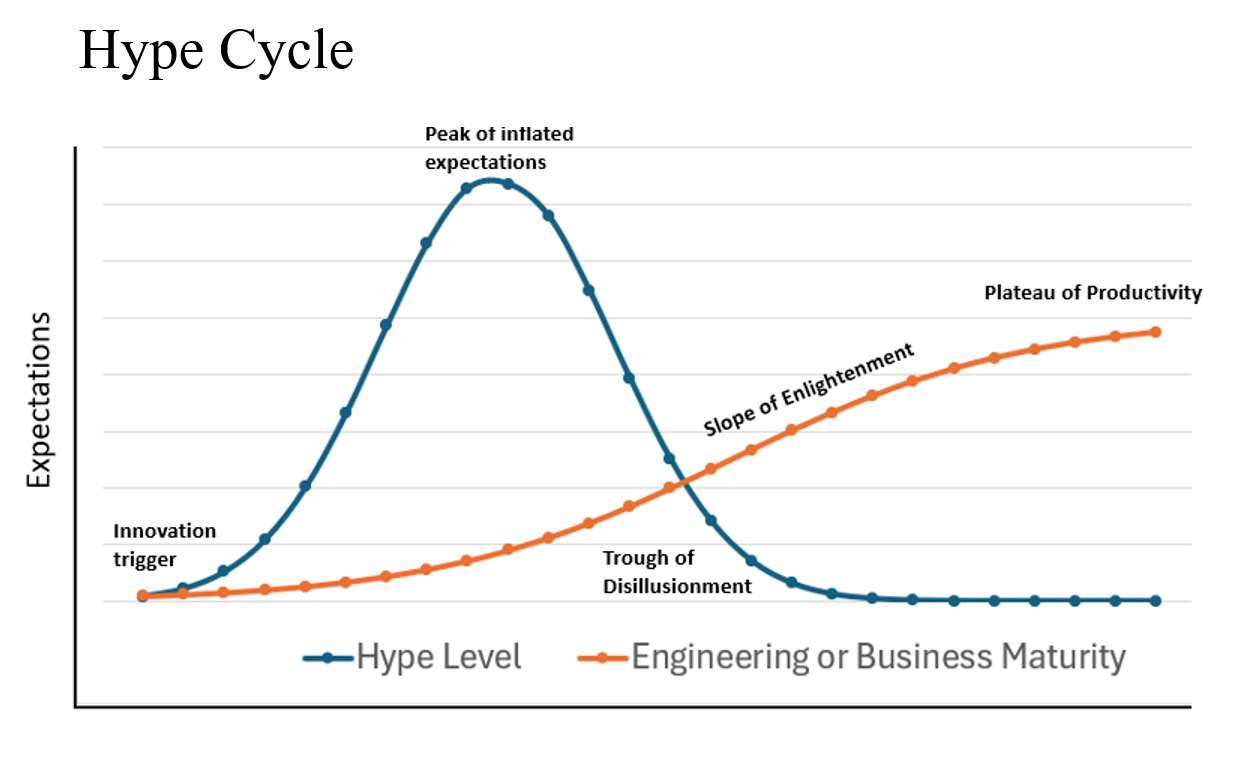}
	\caption{Hype Cycle: Hype Level, Engineering or Businessa Maturity and combination Hype Cycle \cite{steinert2010scrutinizing}}
\label{fig:hype-cycles-brakedown}
\end{figure}

The hype cycle model, introduced by Gartner Inc. in 1995, 
\cite{steinert2010scrutinizing} illustrates the evolution of technological expectations over time. As shown in Figure~\ref{fig:hype-cycles-brakedown} \cite{steinert2010scrutinizing}, the model is formed by merging two distinct curves: a human-centric expectation curve, represented by the dashed blue line, and a classical technology maturity S-curve, represented by the dotted green line. The first equation describes the hype level, which follows a bell-shaped curve driven by excitement, social contagion, and heuristic decision making. This curve captures the tendency of people to overestimate the potential of a technology, leading to an initial peak of inflated expectations, followed by a sharp decline when real-world limitations become evident.

The second equation, the S-curve of technological maturity, represents the gradual advancement of a technology over time. Initially, progress is slow due to limited understanding and minimal performance gains from early investments. However, as knowledge accumulates and improvements accelerate, the technology experiences rapid development until it eventually stabilizes at a plateau defined by its inherent limitations. Together, these two curves form the hype cycle (solid gray line), which models the typical trajectory of emerging technologies, highlighting both the initial overhype and the eventual stabilization at a realistic level of maturity. The stages outlined in the hype cycle depicted in Figure \ref{fig:hype-cycles-brakedown} 
are: 

\begin{enumerate}
    \item \textbf{Innovation Trigger:} A new technology or concept emerges, generating early excitement and media interest, although practical applications are not yet fully understood.
    
    \item \textbf{Peak of Inflated Expectations:} Media hype and early success stories fuel high expectations, leading to overenthusiasm and unrealistic predictions about the technology's potential, with many projects launched but most failing to deliver.
    
    \item \textbf{Trough of Disillusionment:} Initial excitement fades as challenges and limitations become evident, causing skepticism due to failed projects and unmet expectations, leading to a decline in interest and reduced investments.
    
    \item \textbf{Slope of Enlightenment:} A gradual understanding of the technology’s real potential and limitations develops, leading to practical applications and increasing adoption as early adopters recognize value in realistic use cases.
    
    \item \textbf{Plateau of Productivity:} The technology becomes mainstream and widely adopted, with stable, practical use cases delivering real benefits, achieving sustainable growth and a proven track record.
\end{enumerate}

Although the hype-cycle is typically illustrated as a single expectation curve, it actually overlays the S-curve of technology diffusion; how these two curves interact determines both the tempo and the amplitude of successive boom-and-bust phases, yet this relationship remains only sketchily articulated in the literature. Consequently, existing research offers limited guidance on when firms and investors should initiate, scale, or exit their involvement in order to capture upside while containing downside risk.  
Closing this gap will require analytical frameworks that fuse expectation dynamics with adoption kinetics and strategic positioning along the hype-cycle.

\subsection{Review and Synthesis of AI Winters and AI Summers} \label{sec:ai-history-synthesis}

In this section, we synthesise evidence from peer-reviewed papers, technical blogs, and public presentations to trace the recurring boom-and-bust pattern (AI Summers and AI Winters) that has characterised artificial intelligence research and commercialisation. It is important to note that \textbf{AI is not a single invention, but a shifting portfolio of techniques that is constantly redefined over time} — from symbolic reasoning and expert systems to machine learning, deep learning, and today’s large-scale generative models. Each element in this portfolio follows its own trajectory, so the broad history of AI is best understood as the superposition of many overlapping hype cycles rather than one monolithic wave.

These technology-specific curves explain why the field repeatedly climbs to a \emph{Peak of Inflated Expectations}, descends into a \emph{Trough of Disillusionment}, and, where genuine utility is present, climbs the \emph{Slope of Enlightenment} to a durable \emph{Plateau of Productivity}. Recognising AI as a constellation of evolving technologies helps clarify why optimism and disappointment can coexist—and why practical value often emerges only after multiple passes through the cycle.

Take \emph{neural networks} and \emph{software agents} as illustrative examples. For a long time, neural networks were not considered mainstream AI \cite{hinton2023interview}; however, especially during the past 15 years, that has changed. Similarly, introduced in the early 1990s, agent technology ascended to widespread media enthusiasm, encountered technical and market barriers, resurfaced with the rise of web services, and is now experiencing renewed interest through autonomous AI assistants. In other words, the same underlying concept has already traversed the Gartner hype cycle several times.

All available AI history illustrations were analyzed, as listed below:
\cite{menzies200321st,kautz2022third,noguchi2018practical,toosi2021brief,francesconi2022winter,colliot2023non,harguess2022next,perspectives2020,labelf2022,brightworkresearch2020,winiger2017,altmann2023historyAI,latta2021aihistory,schuchmann2019aiwinter,milton2018,sole2022aihstory}.

\begin{figure}[!h]
    \centering
\includegraphics[width=1.0\textwidth]{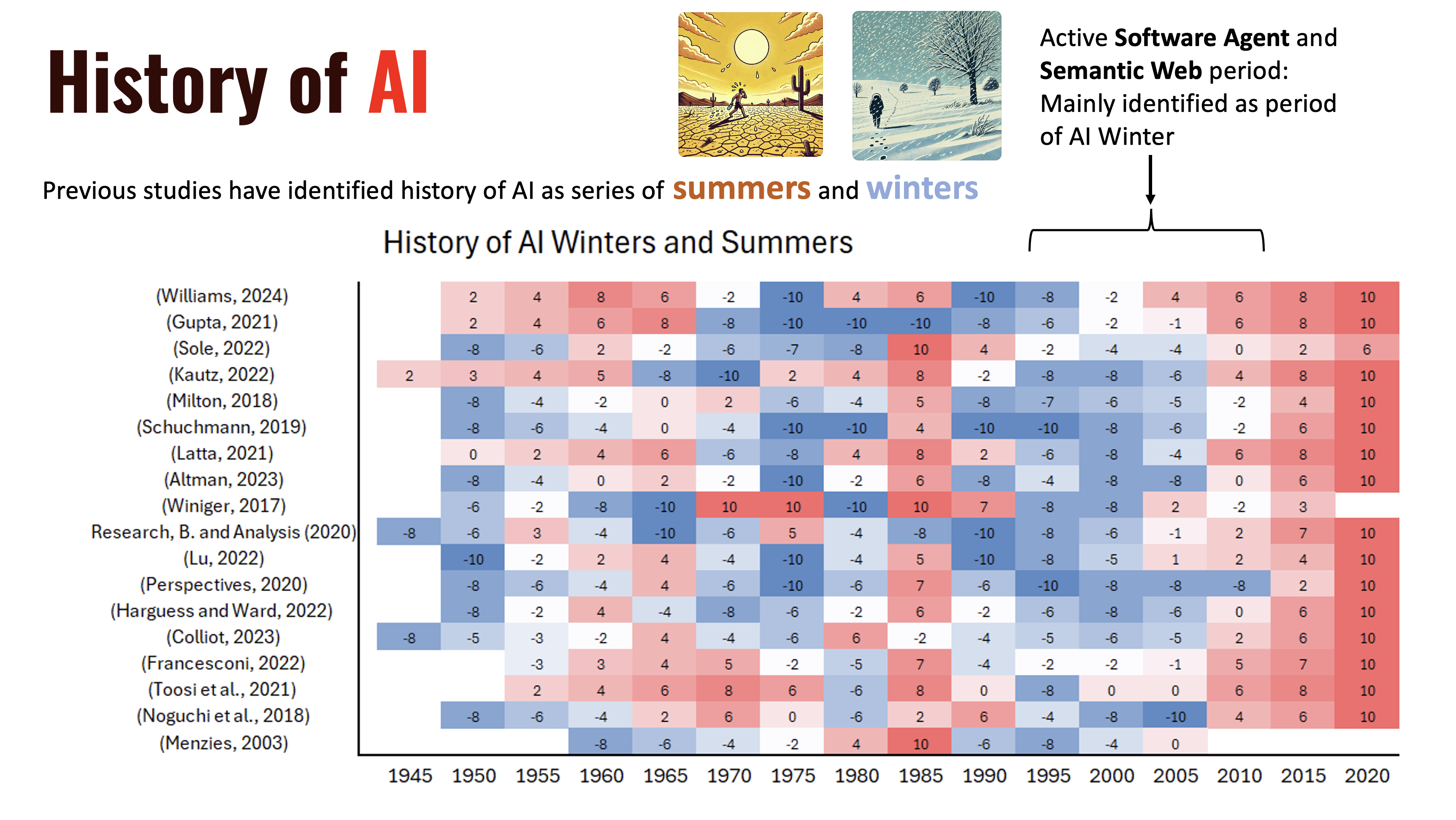}
	\caption{History of Artificial Intelligence as illustrated by previous studies. Period of active Semantic Web development is identified typically as AI Winter period}
\label{fig:ai-history-summary-semantic-web}
\end{figure}

To quantify ``summer'' and ``winter'', a grading scale from +10 to -10 was used, normalizing each illustration to this same scale. This means that all figures receive a maximum value of +10 when they visually reach their peak and -10 when they visually reach their lowest point. This grading serves as an indication of ``temperature,'' verbally expressed as ``AI summer'' (hot) or ``AI winter'' (cold). The aggregated illustration is presented in Figure~\ref{fig:ai-history-summary-semantic-web}.
This figure also indicates that the period of active Semantic Web research is generally considered an AI winter. Furthermore, none of the studies mention either the topic \textit{Semantic Web} or \textit{Software Agents} as representing any form of AI boom in historical AI illustrations.

\subsection{Summary of AI Illustrations}

AI history illustrations known to the authors of this study depict the period from 1995 to 2010 as an AI winter rather than a time of substantial research advancements. These accounts consistently emphasize stagnation and minimal innovation, portraying this period as one characterized by diminished funding, waning public interest, and a general lack of transformative breakthroughs in AI. While some research activities continued during this time, the prevailing perception is that AI did not experience any major surge or revolution.

Furthermore, none of the AI history illustrations reviewed in this study explicitly mention Software Agents or the Semantic Web. This absence suggests that these topics were not widely recognized or highlighted as pivotal elements within the broader narrative of AI development during this period. Their exclusion reinforces the notion that this era was marked more by stagnation than by the emergence of significant new paradigms or technologies.


\section{The Semantic Web and Its Vision of AI Software Agents} \label{sec:semantic-web-agents}

\subsection{Hype Cycles: Software Agent Technology}
Software agent technology has gone through a hype cycle a couple of times during its history. The following quote is from 1996:

\begin{quote}
"Agents are here to stay, not least because of their diversity, their wide range of applicability and the broad spectrum of companies investing in them. As we move further and further into the information age, any information-based organization which does not invest in agent technology 
may be committing commercial hara-kiri." ~\cite{nwana1996software}
\end{quote}

Authors of the above quote wrote another article a few years later ~\cite{nwana:1999} stating that the assumptions were too high. If anything, companies that invested in agent technology in the 1990s were perhaps the ones to commit a form of hara-kiri.

Recently, in December 2024, Microsoft CEO Satya Nadella stated that AI-driven software agents are fully revolutionizing business-to-business (B2B) applications. According to Nadella, AI-driven agents will take over business logic and enable new AI-powered applications, placing software agents at the heart of the current AI hype. The full quote from an episode of the BG2 Pod series, where Nadella stated in an interview video \cite{bg2pod2025}, e.g., is:


\begin{quote}
“I think that the notion that business applications (B2B Apps) exist, that is probably where they all collapse, in the (Software) Agent-era.” -- “Because if you think about it, they (B2B Apps) are essentially CRUD-databases with a bunch of business logic. The business logic will go to these Agents. These Agents are going to be multi-repo-CRUD so they are not going to discriminate between what the backend is. They are going to be updating multiple databases. All the (business) logic will be in the AI-tier (Agent-layer) so to speak.”
“Once the AI-tier becomes the place where all the logic is, then people will start replacing the backends. -- As we speak, I think we are seeing quite high rates of wins in our Dynamics backends and the Agent use. We will go very aggressive trying to collapse it all. Whether it is customer service, be it finance and operations…” -- “People want more AI-native business apps. It means the logic tier can be orchestrated by AI and AI Agents in a very seamless way.”
\end{quote}

Obviously, noting that agent technology has had inflated expectations in the past does not prove anything about the future. This time, it might be different, as the overall technology infrastructure has evolved significantly since the 1990s. We also have to remember that, for example, neural networks have experienced several hype cycles before the major breakthrough in the 2010s.

\subsection{Software Agents Before the Semantic Web}

Long before the Web era, the conceptual foundations for software agents was established by research in \emph{Distributed Artificial Intelligence} (DAI) and the \emph{Actor model} \cite{hewitt1977viewing}.  DAI cast intelligence as an \emph{emergent} phenomenon arising from multiple, loosely-coupled problem solvers that cooperate across a network, thereby elevating autonomy and negotiation to primary design concerns \cite{bond2014readings,gasser2014distributed}.  In parallel, the Actor model supplied a concrete computational metaphor of independent, concurrently executing entities that communicate solely through asynchronous message passing.  Together these two traditions contributed the twin conceptual pillars of \textbf{distribution} and \textbf{concurrency}, which later researchers distilled into the now-standard agent properties of social ability and proactive–reactive behaviour \cite{wooldridge2009introduction}.

The 1990s witnessed the emergence of software agents as a concept in computer science, driven by the need for autonomous and intelligent systems in increasingly complex environments. This decade saw the consolidation of core agent characteristics, including autonomy, proactivity, reactivity, and social ability \cite{nwana1996software,wooldridge1995intelligent}. The rapid expansion of the World Wide Web around 1994 underscored the potential for agents to operate within this new digital space \cite{maes1994agents}. Foundational research explored diverse aspects of agent technology, from theoretical underpinnings and formal methods to practical applications for information management and personalized assistance \cite{wooldridge1995intelligent,maes1994agents}.

The development of various agent architectures, such as reactive, deliberative, and hybrid models, further illustrated the field's commitment to creating sophisticated autonomous entities \cite{weiss1999multiagent}. Researchers also focused on defining the boundaries of agency, distinguishing agents from conventional programs through taxonomies and formal definitions \cite{franklin1996agent}. The intellectual roots of software agents were traced back to Distributed Artificial Intelligence (DAI) and Carl Hewitt's Actor model from 1977, highlighting the long-standing interest in distributed and interacting intelligent systems \cite{nwana1996software,hewitt1977viewing,wooldridge2009introduction}. 

\paragraph{From Definitions to Standards.}%
The explicit taxonomies and formal definitions produced during this period played a pivotal role in \emph{standardising} the concept of agency.  By establishing a common vocabulary, they enabled the Foundation for Intelligent Physical Agents (FIPA) to publish its initial specifications—most notably the \emph{FIPA Agent Communication Language} (FIPA ACL) and a reference architecture that has since underpinned almost every industrial-grade multi-agent platform \cite{fipa1997spec}.  These efforts, together with subsequent W3C initiatives (e.g.\ the \textit{Agent Markup Language} submission), ensured that later Semantic-Web agents could interoperate across organisational and domain boundaries on the basis of well-defined semantics rather than ad-hoc conventions.


\subsubsection{Impact of the World Wide Web’s Expansion on Software Agents}

The exponential expansion of the World Wide Web in the mid-1990s profoundly altered the landscape in which software agents were conceived and applied. Originally developed in the context of distributed AI and local computation environments, software agents found new significance as the Web became a vast, decentralized, and heterogeneous information space. The growing complexity of navigating, searching, and processing this information catalyzed the demand for autonomous systems capable of filtering, retrieving, and reasoning over web-based content on behalf of users.

Seminal work by Pattie Maes and others \cite{maes1994agents} recognized this shift early, introducing agents as personalized assistants that could reduce information overload in the emerging Web era. This period also saw the rise of "information agents"—autonomous programs designed specifically to operate over Web protocols, harvest content, and interact with web services. The scalability and openness of the Web provided a compelling environment for agents to demonstrate their core properties: autonomy, proactivity, and social interaction, especially as HTML and HTTP standards facilitated agent-based crawling and retrieval.

Moreover, the ubiquity of browsers and online services created a new user expectation: software should act intelligently in navigating and managing web interactions. This sociotechnical transformation fueled both academic research and industrial exploration into agent-based architectures, eventually motivating the integration of agents into the foundational vision of the Semantic Web \cite{berners_lee_2001}. Thus, the Web’s growth did not merely provide a backdrop—it actively shaped the functionality, relevance, and perceived utility of software agents during this foundational decade.

\subsection{The Vision of the Semantic Web and Intelligent Agents}

Tim Berners-Lee, James Hendler, and Ora Lassila, in their seminal 2001 paper ``The Semantic Web'' \cite{berners_lee_2001}, envisioned an evolution of the World Wide Web into a more intelligent and interconnected system.  This Semantic Web would be an infrastructure where information has explicitly defined meaning, facilitating human-computer cooperation and enabling machines to understand and process web data through a universal framework of metadata and ontologies. This involved enriching web resources with semantic markup using languages like RDF and OWL, making content machine-processable \cite{mcguinness2004owl}.  The goal was a transition from unstructured documents to interconnected data, enhancing knowledge sharing, reuse, and integration \cite{shadbolt2006semantic}, aligning directly with AI's goals of machine understanding and reasoning \cite{berners_lee_2001}.  As of the time of writing, this highly influential paper has garnered over 30 000 citations.

Intelligent agents were conceived as crucial actors within this vision, able to automatically access, interpret, and reason about the semantically marked-up information \cite{hendler2001agents}.  These agents would act on behalf of users, performing tasks such as information retrieval, service discovery, automated decision-making (e.g., scheduling meetings or making purchases), and seamless interaction across different platforms \cite{mcilraith2001semantic}.  Personal assistants, guided by rules, ontologies, and user profiles, were highlighted as prime examples of potential applications \cite{lassila2001enabling}.

Realising this vision depends on the \emph{ability of autonomous agents to converse}.  The FIPA ACL specification meets this requirement by supplying a performative-based message envelope whose semantics are defined in a modal logic of communicative acts \cite{fipa2002acl}.  When the \texttt{content} field of such messages is expressed in RDF or OWL, the resulting utterances are simultaneously interpretable at the \emph{data} and the \emph{speech-act} levels, thereby enabling richly contextualised dialogues among heterogeneous Web agents \cite{vreeswijk2005dialogue}.  Accordingly, ACLs form the critical bridge between the Semantic Web’s data layer and the multi-agent community’s interaction layer.

The paper's emphasis on AI Software Agents, and the backgrounds of Hendler and Lassila in intelligent agents and knowledge representation, underscore the Semantic Web's connection to advancing autonomous systems capable of reasoning, decision-making, and communication \cite{berners_lee_2001}. By leveraging standardized semantics, agents could interpret and act upon information with greater sophistication.  This vision linked the future of the Semantic Web to developments in AI and positioned Software Agents as a critical component in realizing the Web's full potential.  While the full vision remains unrealized, it remains a cornerstone in the intersection of Semantic Web technologies and AI Software Agent development.


\subsection{Key Enabling Technologies and the Semantic Web Stack}

Several technologies were crucial for realizing the vision of the Semantic Web and its integration with intelligent agents.  The Resource Description Framework (RDF) provided a standard model for data interchange on the web, enabling the expression of relationships between resources \cite{lassila1998resource}.  The Web Ontology Language (OWL), built upon RDF, was designed to represent rich and complex knowledge about entities, their classifications, and relationships \cite{mcguinness2004owl,horrocks2003owl}. Ontologies, formal and explicit specifications of shared conceptualizations, provided a common vocabulary and understanding of specific domains, which was crucial for enabling agents to interpret information consistently \cite{gruber1993toward}.

Agent Communication Languages (ACLs), such as the Foundation for Intelligent Physical Agents (FIPA) ACL \cite{fipa2002fipa}, were recognized as essential for facilitating interaction and information exchange between autonomous agents.  Furthermore, the \emph{DAML-S} ontology—renamed \emph{OWL-S} after the release of OWL—\cite{martin2004owl} was created to add a machine-interpretable semantic layer to conventional Web-service descriptions. It exposes three mutually supportive components: a \emph{Service Profile} that advertises what the service does, a \emph{Process Model} that specifies how the service can be executed, and a \emph{Grounding} that binds abstract inputs and outputs to concrete message exchanges. Together these components allow an intelligent agent to match its goals to advertised profiles and thus discover relevant services, to use the process model and grounding information to construct and send the correct messages for direct invocation, and to chain multiple process models into coherent workflows for automatic composition of complex tasks. The development of DAML-S/OWL-S therefore represents a focused effort to supply the semantic infrastructure—rooted in ontologies, RDF, and OWL—along with the inter-agent communication mechanisms required for truly autonomous agents on the Semantic Web.

\begin{figure}[!h]
    \centering
    \includegraphics[width=0.5\textwidth]{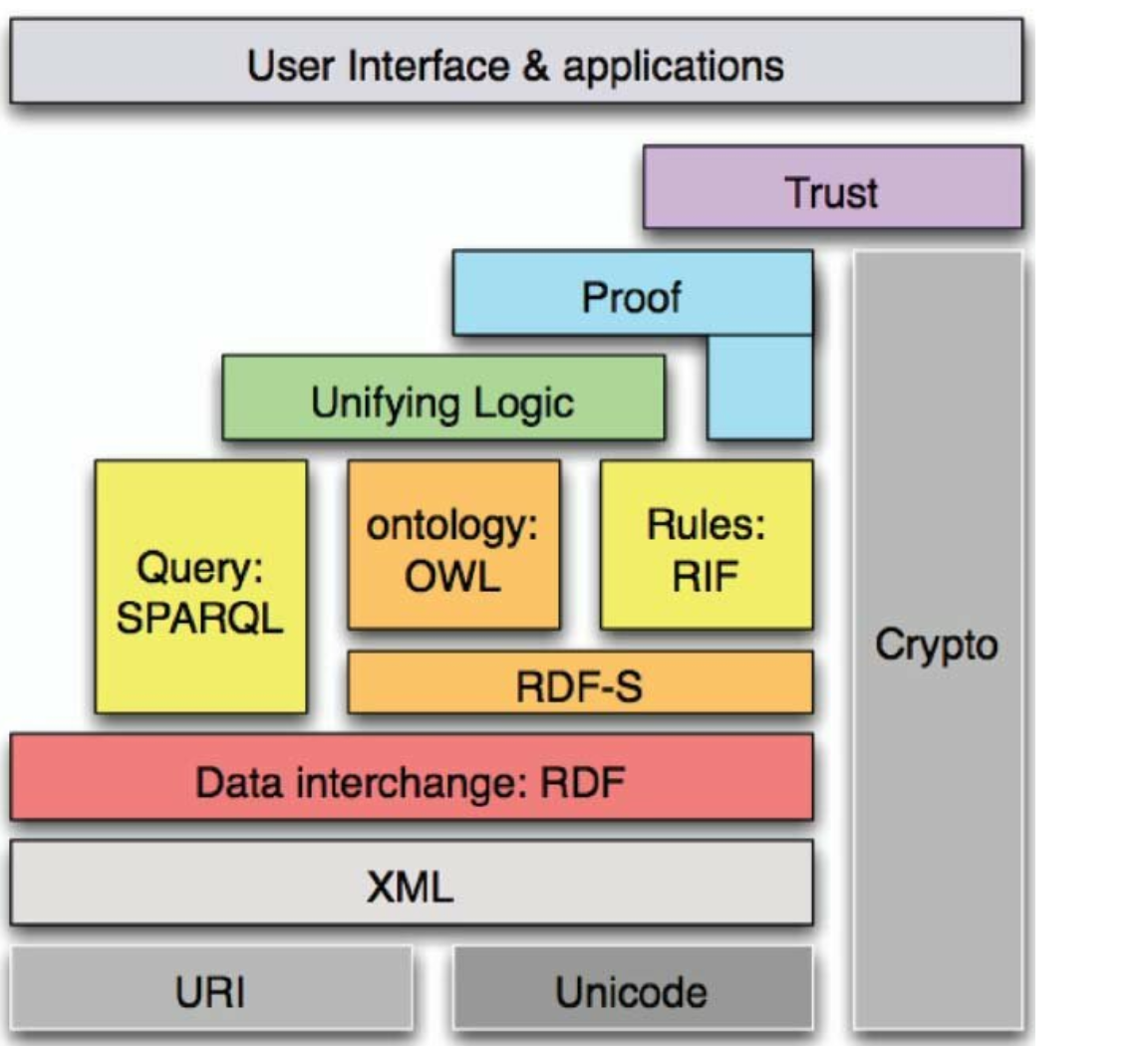}
    \caption{Semantic Web Stack by Tim Berners-Lee 2006 \cite{semanticWebstack2006}}
    \label{fig:semantic-web-stack}
\end{figure}

Figure~\ref{fig:semantic-web-stack} illustrates the "Semantic Web Stack," a layered architecture proposed by Tim Berners-Lee for organizing the technologies of machine-readable web data \cite{semanticWebstack2006}.  URIs and Unicode provide the foundation for identification and character encoding.  XML offers a syntax for structured documents, while RDF and RDF Schema (RDFS) define a standardized way to describe resources and their relationships \cite{brickley2004rdf}.  OWL builds on this to enable the creation of complex ontologies. Rule languages like RIF (Rule Interchange Format) and SWRL (Semantic Web Rule Language), along with the SPARQL query language, were intended to support inference and querying over this semantically rich data \cite{horrocks2004swrl}. 

While OWL enables inference through description logic constructs such as subclassing, property hierarchies, and transitive closures, its expressivity is intentionally constrained to ensure decidability and tractable reasoning. To address use cases that require more expressive logic—such as conditional rules and inter-property dependencies—rule-based extensions were introduced. The Semantic Web Rule Language (SWRL) \cite{horrocks2004swrl} allows the combination of OWL ontologies with Horn-like rules of the form \textit{if antecedent then consequent}, significantly enhancing inferencing capabilities. For example, SWRL can express rules like "if a person has a parent who is female, then the parent is a moth

Higher layers of the stack were envisioned to handle logical reasoning, proof verification, and trust establishment through cryptographic methods, ultimately enabling applications with enhanced interoperability. Over the past two decades, the Semantic Web community has converged on a set of concrete cryptographic primitives that operationalise the previously aspirational \emph{proof} and \emph{trust} layers.  A prerequisite for any digital signature is a deterministic byte-level representation of the graph being protected; this is now provided by the
\emph{RDF Dataset Canonicalisation and Hash 1.0} (RDFC-1.0) algorithm, which yields a stable hash despite blank-node renaming and graph isomorphisms~\cite{w3c_rdf_canon_2024}. Once canonicalised, graphs can be enveloped by the \emph{Data Integrity 1.0} suite (formerly “Linked-Data Signatures”), embedding cryptographic proofs directly in RDF or JSON-LD and supporting selective-disclosure cryptosuites such as BBS+~\cite{w3c_data_integrity_2025}.

Building on these primitives, the \emph{Verifiable Credentials Data Model 2.0} binds claims to subjects in tamper-evident envelopes that agents can verify offline, while \emph{Decentralized Identifiers (DID) Core} replaces hierarchical certificate authorities with
URL-resolvable public-key documents~\cite{w3c_vc_2_0_2025,w3c_did_core_2022}.  
Together these technologies allow autonomous agents—whether classic rule engines or modern LLM pipelines—to (i) verify the provenance of data before reasoning, (ii) attach signed proofs to their
own outputs, and (iii) assemble verifiable chains of custody across heterogeneous knowledge graphs.

\subsubsection{Semantic Web Rollout Vision}

\begin{figure}[!h]
    \centering
    \includegraphics[width=0.75\textwidth]{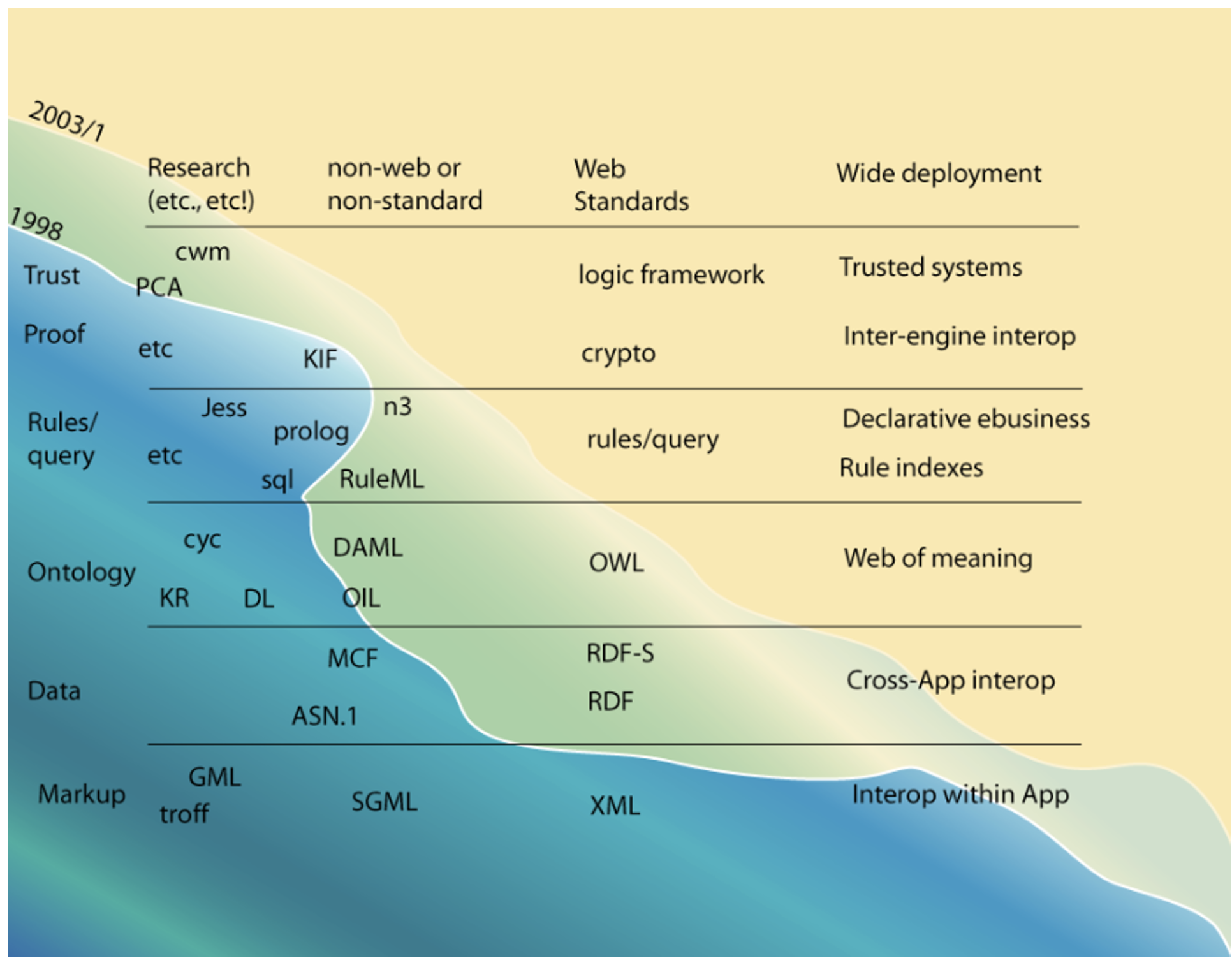}
    \caption{The Semantic Web Wave 2003 by Tim Berners-Lee \cite{BernersLee2003SemanticWebWave}}
    \label{fig:semantic-web-wave}
\end{figure}

Figure~\ref{fig:semantic-web-wave} depicts Tim Berners-Lee's 2003 "rollout vision" for the Semantic Web, projecting the integration of layered technologies from basic markup (SGML, XML, RDF) to advanced constructs (OWL, DAML+OIL) and logic (Prolog, RuleML), culminating in proof and trust mechanisms for automated reasoning and secure interoperability \cite{BernersLee2003SemanticWebWave}.

However, this vision was not fully realized.  While RDF and OWL saw adoption in niche areas, widespread uptake was hampered by competing technologies and shifting industry priorities.  The higher layers of the Semantic Web stack, particularly those based on formal logics, faced limited adoption due to the complexity of technologies like OWL dialects and fundamental differences in assumptions compared to traditional database systems. OWL's Open World Assumption (unknown facts are not presumed false) and lack of a Unique Names Assumption (identical names don't necessarily mean identical entities) contrasted with the closed-world and unique-names assumptions prevalent in many database applications.  Despite initiatives like the W3C's Semantic Web Activity program and efforts on Web Services (e.g., OWL-S) to advance intelligent software agents as envisioned in the original Semantic Web paper \cite{berners_lee_2001}, this direction did not achieve the expected success within the Semantic Web context.

\subsection{Semantic Web Influence on Modern AI/ML Data Integration}

Although the Semantic Web’s most ambitious goals remain aspirational, its foundational interoperability standards—such as RDF, OWL, SPARQL, and SHACL—have significantly influenced how structured data is integrated in modern AI and machine learning systems. These standards provided early blueprints for achieving semantic consistency across heterogeneous datasets, a challenge that remains central to AI pipelines today.

In large-scale AI systems, knowledge graphs—many of which adopt RDF or OWL-based schemas—serve as key instruments for data integration, enabling entity resolution, feature alignment, and graph-based feature engineering. Companies like Google, Amazon, and Microsoft have adopted internal ontologies to normalize concepts across business domains, facilitating model interoperability and multi-source training \cite{shadbolt2006semantic}. SPARQL’s graph-query paradigm also anticipates the emerging need for expressive retrieval interfaces in retrieval-augmented generation (RAG) systems.

Furthermore, recent approaches to embedding knowledge graphs into vector spaces for use in ML models draw on ontology axioms to constrain embedding learning, enabling logic-conformant representations \cite{nickel2016review}. Validation languages such as SHACL have also inspired schema-checking layers in AI pipelines, ensuring that training and inference data adhere to expected semantic contracts. In effect, these practices embed Semantic Web–style interoperability principles at the foundation of modern machine learning workflows.

Thus, although seldom cited by name, the Semantic Web’s influence persists through the design patterns, vocabularies, and validation logic that now underpin scalable and trustworthy AI data infrastructures.

\subsection{Software Agents in the LLM Age 2017 onwards} \label{sec:software_agents_are_hot}

The advent of large language models (LLMs) after the advent of the Transformer \cite{vaswani2017attention} has dramatically reshaped the landscape of software agent development, opening up new possibilities and challenges.  Pre-LLM, agents often relied on meticulously crafted rules, ontologies, and knowledge bases, limiting their adaptability and generalizability \cite{wooldridge2009introduction,russell2010artificial}.  While these approaches provided a degree of control and explainability, they struggled to handle the complexity and ambiguity of real-world scenarios. LLMs, trained on massive datasets, offer a fundamentally different approach.  They possess an unprecedented ability to understand and generate natural language, reason, and even exhibit some degree of common sense, albeit imperfectly \cite{brown2020language,ouyang2022training,wei2022chain}. This inherent capability allows LLM-powered agents to interact with users and environments in a much more natural and flexible way, reducing the need for extensive manual programming and domain-specific knowledge engineering.

The integration of LLMs with agent frameworks has led to the emergence of "LLM-powered agents" or "LLM agents," capable of performing a wider range of tasks than their predecessors. The simple definition for an LLM Agent is an artificial entity with prompt specifications (initial state), conversation trace (state), and ability to interact with the environments such as tool usage (action) \cite{cemri2025multiagentllmsystemsfail,naveed2024comprehensiveoverviewlargelanguage}. These agents can leverage LLMs for various functions, including natural language understanding (NLU), natural language generation (NLG), planning, decision-making, and even tool use \cite{xi2023rise,wang2023survey,yang2023foundation}.  For instance, an LLM agent can be prompted to break down a complex task into smaller, manageable sub-tasks, search the web for relevant information, interact with APIs to execute actions, and summarize the results for the user \cite{yao2022react,shinn2023reflexion}. Frameworks like LangChain \cite{chase2022langchain} and AutoGen \cite{wu2023autogen} provide tools and abstractions to facilitate the development and deployment of these agents, enabling developers to chain together LLM calls, external tools, and custom logic. This marks a significant shift from symbolic AI to a more data-driven, emergent approach to agent development.

Despite the significant advancements, LLM agents also present unique challenges.  One major concern is the potential for "hallucinations," where the LLM generates plausible-sounding but factually incorrect or nonsensical information \cite{ji2023survey}.  This can lead to unreliable agent behavior and requires careful mitigation strategies, such as grounding the LLM's responses in external knowledge sources or employing techniques like retrieval-augmented generation (RAG) \cite{lewis2020retrieval}. Other challenges include ensuring the safety and ethical behavior of agents, managing the computational cost of running large models, and addressing biases inherent in the training data \cite{bender2021dangers,weidinger2021ethical}. Furthermore, the "black box" nature of LLMs can make it difficult to understand the reasoning behind an agent's actions, hindering debugging and limiting trust in critical applications \cite{doshi2017towards}.

The ongoing research in LLM-powered agents focuses on addressing these limitations and exploring new capabilities with the emerging multi-agent systems (MAS).  This includes developing methods for improving the factual accuracy and reliability of LLMs, designing more robust and explainable agent architectures, and creating effective evaluation metrics to assess agent performance across diverse tasks \cite{nakano2021webgpt,ram2023context}.  The convergence of LLMs and software agents represents a significant step towards building more general-purpose, adaptable, and intelligent systems, with the potential to transform how we interact with computers and automate complex tasks. However, responsible development and deployment, with a keen awareness of the inherent challenges, are crucial to realizing the full potential of this technology.

\subsection{Prominent LLM Agent Frameworks and Libraries}

A fast-maturing ecosystem of open-source frameworks now captures the canonical design patterns for
agentic large-language-model (LLM) systems, letting researchers focus on reasoning strategies rather
than boiler-plate glue code.  \textbf{LangChain} supplies composable primitives—tools, memories and
chains—that underpin many production chatbots, document-question-answering (QA) pipelines and research
assistants.  Building on the same abstractions, \textbf{LangGraph} introduces a graph-based execution
engine with explicit state, branching and parallelism, which is invaluable for long-horizon planning
and multi-agent coordination.

Microsoft’s \textbf{AutoGen} formalises ``chat-loop’’ interactions in which several LLMs, optionally
augmented by external tools or human feedback, iteratively critique and refine one another to solve
non-trivial tasks.  Where private or proprietary data are involved, \textbf{LlamaIndex} offers a
complete data layer—index construction, retrieval interfaces and knowledge-graph support—for
retrieval-augmented generation (RAG) over local corpora.  \textbf{CrewAI} targets role-oriented
workflows: agents endowed with complementary expertise share a common memory to tackle coordinated
problem solving or simulation environments. For production use, Microsoft’s \textbf{Semantic Kernel} provides a language-agnostic plug-in system,
a symbolic planner and robust security controls that simplify the integration of generative AI into
enterprise software stacks.  Google’s \textbf{Agent Development Kit} (ADK) implements the
Agent-to-Agent (A2A) protocol and offers orchestration, safety guards and evaluation utilities for
multi-agent research at scale.  Also from Google, \textbf{Genkit} delivers a code-first, TypeScript
workflow with flow orchestration, prompt/version control (\texttt{.prompt} files), observability and
one-command Cloud Run deployment.  Finally, \textbf{Marvin} provides a pythonic decorator-based API
that turns ordinary functions into threaded, asynchronous ``AI functions,’’ making rapid prototyping
inside data-science or machine-learning pipelines straightforward. Additional libraries—including Haystack, 
Embedchain, SuperAGI, Dify and OpenAgents—continue to expand the design space and keep the agent-framework 
landscape highly dynamic.

\subsection{Could Semantic Web Principles Complement LLMs?}
Large-language-model pipelines still lack explicit semantics, provenance, and
verifiability.  Drawing on the Semantic-Web stack, we outline three concrete
integration patterns that harness ontologies and knowledge graphs to mitigate
hallucination and to provide user-facing rationales.

\textbf{Why ontologies?}  
Unlike the high-dimensional token spaces in which large language models (LLMs)
operate, an ontology offers a \emph{canonical, machine-interpretable vocabulary}
of entities, relations and axioms.  Mapping generated spans to URI-identified
concepts therefore gives each symbol an unambiguous referent, providing the
first step toward faithful explanations
\cite{confalonieri2023multiplerolesontologiesexplainable}.

\textbf{Complementary trade-offs.}
OWL and RDF guarantee sound, monotonic inference but require careful manual
modelling, limit expressivity to stay decidable, adopt an Open World Assumption
and incur worst-case exponential reasoning cost.  LLMs sit at the opposite end
of the spectrum: they learn broad coverage automatically, answer in
closed-world style with constant-time decoding, and tolerate noisy data—yet
offer no built-in notion of truth, identity or provenance.  Combining the two
thus pairs the symbolic rigour of OWL/RDF with the statistical reach of LLMs.

\textbf{Pattern 1 – ontology-constrained decoding.}  
During generation the model is prompted or fine-tuned so that every entity
mention must be resolvable to an ontology concept; illegal continuations are
pruned.  This drastically reduces hallucinations and yields provenance for
every token.  A practical example is the plant-disease classifier of
\cite{amara2024enhancingexplainabilitymultimodallarge}, where OWL reasoning is
used online to check that the captioned visual attributes entail a unique
disease class.

\textbf{Pattern 2 – knowledge-graph RAG.}  
Retrieval-augmented generation in which the retriever indexes a knowledge graph
rather than free text grounds answers in triples whose semantics are explicit.
Recent systems such as \textsc{SubgraphRAG} show that even small models achieve
state-of-the-art accuracy \emph{and} produce subgraph-level rationales that
users can inspect \cite{li2025simpleeffectiverolesgraphs}.

\textbf{Pattern 3 – neuro-symbolic post-hoc explanation.}  
After an LLM produces an answer, a reasoning engine aligns the answer with the
ontology, derives minimal entailment paths, and verbalises them as
“because-clauses.”  Confalonieri and Guizzardi identify reference modelling,
commonsense reasoning and knowledge-refinement as the three main roles
ontologies play in such explanatory pipelines
\cite{confalonieri2023multiplerolesontologiesexplainable}.

\subsection{How AI Has Shaped Real-World Semantic Web Applications}

Since the publication of the 2001 Semantic Web vision, advances in AI have continually reshaped how Semantic Web technologies are applied in practice. While the original vision emphasized agent-based reasoning over RDF/OWL graphs, real-world adoption accelerated primarily in areas where AI provided scalable solutions for search, classification, and entity resolution.

Notably, the integration of AI with Semantic Web principles led to the widespread adoption of knowledge graphs, now core components of systems like Google’s Knowledge Panel and Facebook’s Entity Graph. These applications use statistical entity linking and machine learning–driven relation extraction, but rely on RDF-style schemas for consistent interpretation. Similarly, the emergence of schema.org as a lightweight ontology for Web content marked a convergence between SEO optimization and semantic markup, driven by AI-based search and ranking algorithms.

In recent years, machine learning pipelines have begun to adopt semantic validation layers—such as SHACL—for ensuring data integrity, especially in regulated domains like healthcare and finance. AI-driven recommendation systems now leverage ontologies to improve explainability and diversity of results, while retrieval-augmented generation (RAG) models increasingly use graph-based backends to ensure factual grounding.

Thus, while the original agent-centric vision remains largely aspirational, AI has incrementally transformed the Semantic Web into a practical substrate for large-scale, machine-interpretable data systems. These hybrid architectures demonstrate that AI advances have not only extended but redefined the scope of Semantic Web applications in real-world systems.




\section{Academic Significance and Statistics} \label{sec:academic-statistics}

In this section, we use Google Scholar and other bibliometric tools to evaluate the significance of the research wave surrounding the Semantic Web. By analyzing citation counts, publication trends, and related metrics, we assess the impact and reach of Semantic Web research within the academic community. This evaluation provides insight into how the Semantic Web concept influenced subsequent studies, interdisciplinary collaborations, and its adoption in both AI and broader technological domains. Such an analysis helps contextualize the Semantic Web's role in shaping research agendas and its interplay with advancements in AI, particularly in the development of intelligent systems and software agents.

\subsection{Was the Semantic Web a Significant Research Wave? - Google Scholar Analysis}

\begin{figure}[!h]
    \centering
\includegraphics[width=0.9\textwidth]{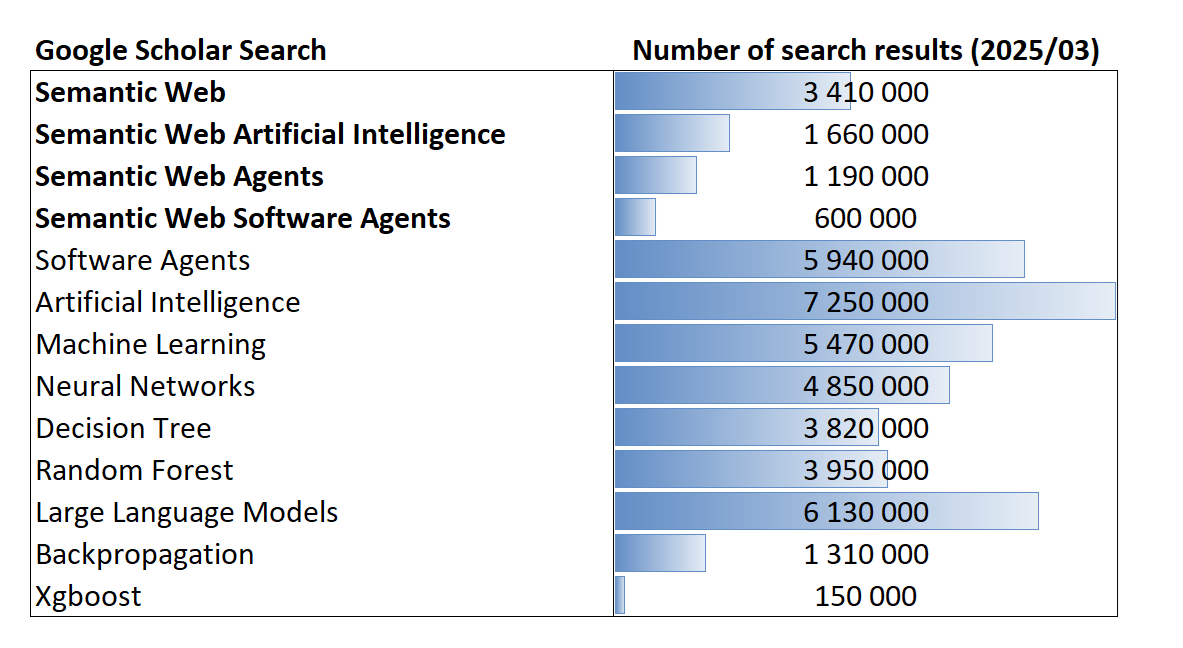}
	\caption{Academic search result hits from Google Scholar using different search terms without time filters}
\label{fig:google-scholar-stats-1}
\end{figure}

To assess the significance of the Semantic Web and its associated AI trends compared to other common AI research themes, we conducted Google Scholar searches, with the results presented in the Table \ref{fig:google-scholar-stats-1}\footnote{Google Scholar–based metrics have well-known caveats:  (i) uneven disciplinary and language coverage (e.g., conference proceedings and non-English venues indexed less consistently than major English journals);  (ii) inclusion of non-peer-reviewed or duplicate records that can inflate counts;  (iii) continuous re-indexing that causes citation numbers to fluctuate over time; and  (iv) keyword matching, where the search string may appear only tangentially in a document, so the retrieved set is not guaranteed to be semantically aligned with the intended topic.  Accordingly, all figures reported here should be read as indicative rather than definitive.}.

The Semantic Web was undoubtedly a high-magnitude research wave, as evidenced by 3.41 million Google Scholar search results (March 2025). While broader topics like ``Artificial Intelligence'' (7.25M) and ``Machine Learning'' (5.47M) have higher volumes, the Semantic Web’s specific focus on structured, meaningful data integration, along with its connection to related areas such as Artificial Intelligence (1.66M) and Agents (1.19M), underscores its significant impact. Despite its narrower scope, the Semantic Web established foundational principles that continue to influence modern AI and data technologies, solidifying its importance in the research landscape.

\subsection{Semantic Web in Google Books}

\begin{figure}[!h]
    \centering
\includegraphics[width=1.0\textwidth]{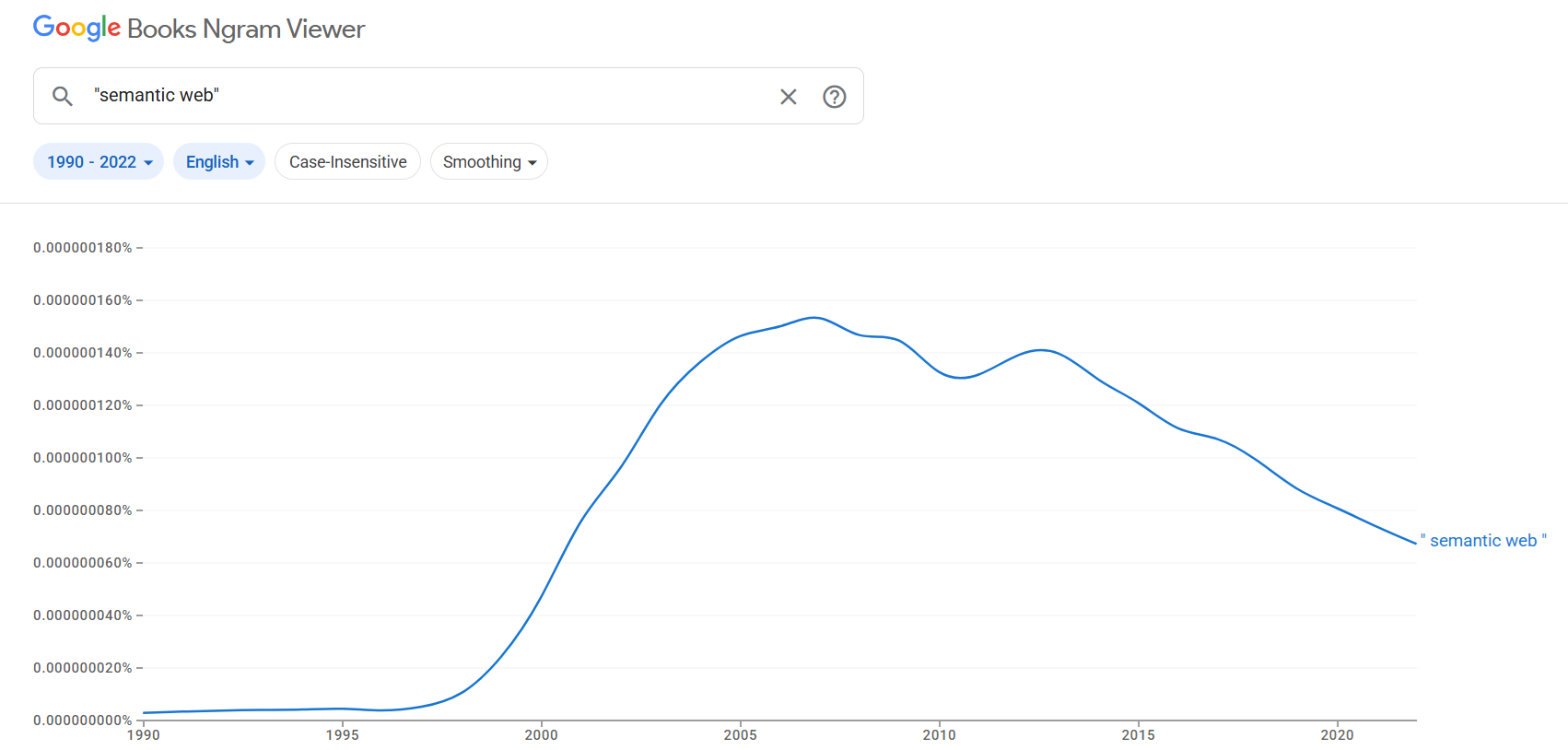}
	\caption{Google Books: number of hits using ``Semantic Web'' as key words (Data not available after 2022)}
\label{fig:google-books-1}
\end{figure}


Figure~\ref{fig:google-books-1} presents a screenshot from the Google Books Ngram Viewer, illustrating the historical prominence of the term Semantic Web in English-language books from 1990 to 2022. The graph shows a significant rise in usage beginning in the early 2000s, coinciding with the publication of the seminal 2001 paper by Tim Berners-Lee, James Hendler, and Ora Lassila, which introduced the concept. Usage peaks between 2005 and 2010, reflecting heightened academic and industrial interest in its potential to transform data structuring and access on the Web. However, the decline after this peak suggests a waning emphasis on the Semantic Web in scholarly and public discourse, possibly due to challenges in realizing its full vision or a shift toward emerging technologies such as machine learning and deep learning. This trend underscores the cyclical nature of technological interest, aligning with broader discussions of the hype cycle.

\subsection{When Did Software Agent Research Emerge?}

\begin{figure}[!h]
    \centering
\includegraphics[width=0.9\textwidth]{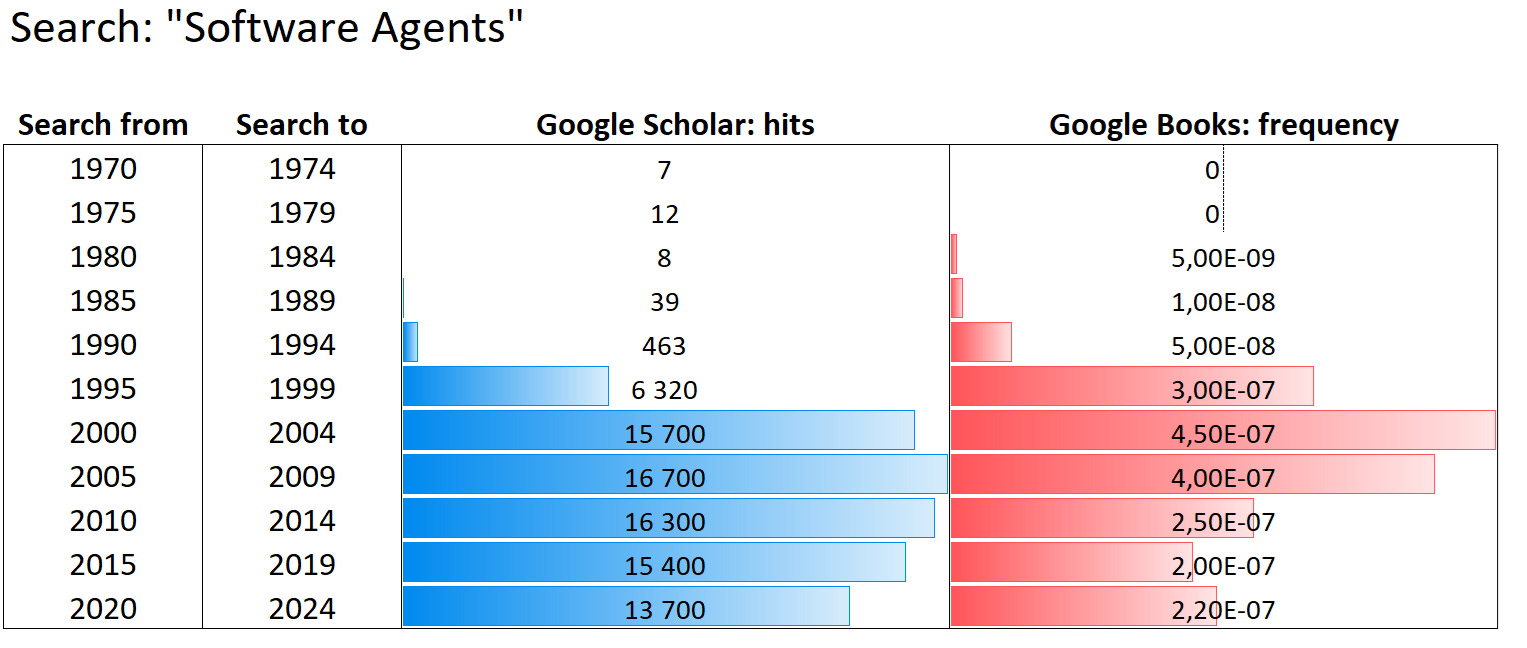}
	\caption{Search results with ``Software Agents'' with five year time intervals Google Scholar (hits) and Google Books (frequencies)}
\label{fig:google-scholar-books-agents}
\end{figure}

One of the most promising future approaches is integrating software agents with large language models (LLMs) to enable not only discussions but also the execution of actions and commands by applications. To analyze the origins and activity trends of software agent research, we compiled statistics from Google Scholar and Google Books Ngram Viewer, with the results presented in Figure~\ref{fig:google-scholar-books-agents}.

Google Scholar searches indicate that the term ``Software Agents'' appeared with minimal activity in the 1970s and 1980s, with only a handful of mentions. Significant growth occurred in the 1990s, reaching 463 results between 1990 and 1994. The field experienced exponential growth after 2000, peaking between 2005 and 2015, with over 16,000 search results in each five-year period. Activity has slightly declined in recent years, with 15,400 results from 2015 to 2020 and a projected 13,700 from 2020 to 2024, indicating sustained but slightly reduced interest in the field.




Figure~\ref{fig:google-scholar-books-agents} highlights the historical prominence of the term ``Software Agents'' in English-language books from 1990 to 2019, showing a significant rise in usage between 2000 and 2010. This trend suggests that Software Agent research received substantial attention during this period, coinciding with the heightened focus on the Semantic Web. The overlap is noteworthy, as the original Semantic Web vision in 2001 envisioned a machine-readable web where Software Agents could perform complex tasks on behalf of users. The simultaneous growth in interest in Software Agents and the Semantic Web underscores their interconnectedness, with Software Agents positioned as a key application of Semantic Web technologies. This period represents a peak in academic and industrial enthusiasm for intelligent, autonomous systems capable of leveraging structured web data, aligning with broader trends in artificial intelligence research.

\begin{figure}[!h]
    \centering
\includegraphics[width=1.0\textwidth]{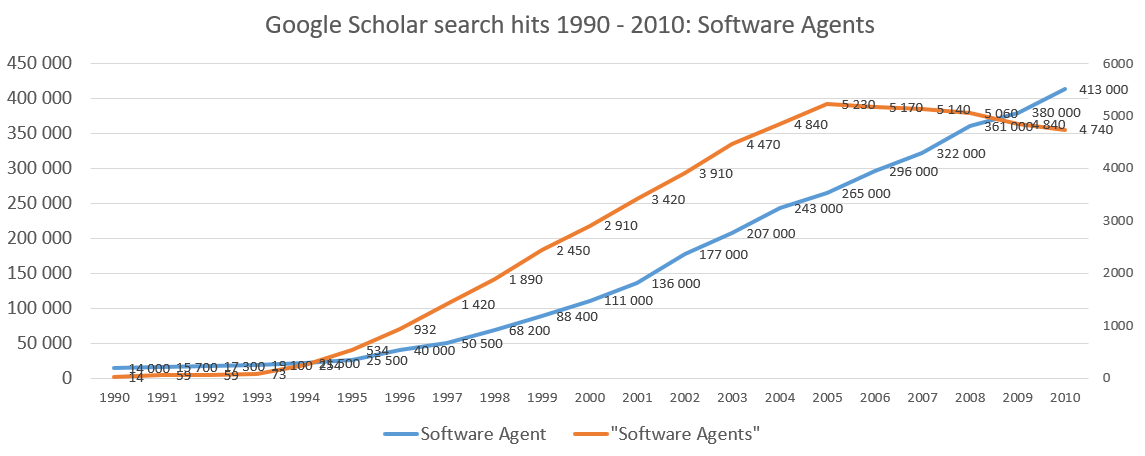}
	\caption{Citation Trends in Software Agent Research: Google Scholar Cites Per Year. Data extracted from Google Scholar.}
\label{fig:software-agent-yearly}
\end{figure}

The illustration in Figure~\ref{fig:software-agent-yearly} depicts the number of Google Scholar search hits for the general terms Software Agent and specific combination "Software Agents" from 1990 to 2010, highlighting their academic prevalence over time. The data indicates a steady increase in publications for both terms, with Software Agent experiencing continuous growth and peaking around 2010 at approximately 413,000 hits. In contrast, the more specific term "Software Agents" saw rapid growth until about 2005, reaching a peak of approximately 5,230 hits before experiencing a slight decline. This suggests that while interest in the broader concept of software agents continued to grow, research using the specific phrase "Software Agents" peaked mid-decade before stabilizing or declining.

\section{Discussion} \label{sec:discussion}

This paper has provided a review and synthesis of previous studies on AI history, particularly the cycles of AI summers and AI winters in Section~\ref{sec:ai-history}. We introduced a normalized temperature scale to quantify these fluctuations. Section~\ref{sec:semantic-web-agents} explored the Semantic Web's foundational vision and its relationship with AI-driven Software Agents, while Section~\ref{sec:academic-statistics} analyzed bibliometric data and academic trends, highlighting the Forgotten AI Wave.

\subsection{Software Agents: A Hot Topic—But Where Did They Come From?}

As discussed in Section~\ref{sec:software_agents_are_hot}, Software Agents are currently a hot topic, with Microsoft’s CEO recently predicting that AI-driven Software Agents will disrupt B2B software applications. 

Where did agents originate? AI history is marked by cycles of optimism and setbacks. In Section~\ref{sec:ai-history}, we reviewed prior studies and illustrations of AI history, concluding that the Semantic Web and Software Agents represent a neglected phase in AI’s evolution. This section explores why this research wave did not gain widespread recognition, the challenges that limited its impact, and how modern AI developments may revive its core principles.

\subsection{Was the Term Artificial Intelligence a Curse?}

A key question arises: Why wasn’t the Semantic Web recognized as a distinct AI wave? As revisited in Section~\ref{sec:semantic-web-agents}, the original Semantic Web vision in 2001 sought to create a machine-readable web enabling intelligent reasoning and automation. Software Agents played a central role, designed to operate autonomously, interact with structured data, and execute user tasks. During the early 2000s, Semantic Web research saw a surge in academic publications and funding, as evidenced by our citation analysis.

One possible explanation, as suggested by Ora Lassila in his keynote talk \cite{lassila2024semantic}, is that Artificial Intelligence was perceived negatively at the time, leading researchers to distance themselves from the term. Instead, they used alternatives terms, such as Software Agents and Knowledge Representation. A similar trend occurred in the early 1980s when Geoffrey Hinton and other researchers avoided using the term Neural Networks due to its controversial status within the AI community \cite{ford2018architects}.

\subsection{Why Didn’t Software Agents on the Semantic Web Succeed?}

Despite initial momentum, the Semantic Web failed to become a mainstream AI paradigm. One primary reason was the dominance of symbolic reasoning and logic-based AI at the time, which contrasted with the statistical and data-driven approaches that later fueled AI’s success. Software Agents lacked access to the powerful natural language processing (NLP) and deep learning techniques that characterize modern AI applications.

Although the Semantic Web did not achieve its grand vision of intelligent software agents, its principles are resurfacing in contemporary AI, particularly through Large Language Models (LLMs) and AI-driven Software Agents.

\subsection{Revisiting the Semantic Web in the Age of LLMs}

Recent advancements highlight the renewed relevance of Semantic Web concepts such as structured knowledge representation, reasoning, and autonomous agents. Knowledge graphs, an evolution of Semantic Web technologies, are widely used by companies like Google and Microsoft to enhance AI-driven search and recommendation systems. Moreover, modern Software Agents are now integrated with LLMs, enabling them to interpret unstructured text, execute commands, and autonomously interact with various data sources—effectively realizing the Semantic Web’s vision through alternative means.

A growing emphasis on AI explainability, data provenance, and structured knowledge integration suggests that Semantic Web technologies could find new applications in the evolving AI landscape. Hybrid neuro-symbolic AI approaches that combine statistical learning with symbolic reasoning may leverage the Semantic Web’s structured data to enhance AI reasoning and interpretability.

\subsubsection{Trustworthy Agents in the LLM Era}
\label{sec:trustworthy-agents}

LLM‐powered agents inherit the Semantic Web’s long‐standing challenge of trust: \emph{how can an autonomous system be certain that the data it consumes or the code it executes is authentic and untampered?}  By leveraging the cryptographic primitives, an agent can (i) verify a Verifiable Credential–signed knowledge‐graph fragment before reasoning~\cite{w3c_vc_2_0_2025}, (ii) attach Data Integrity–based proofs to its chain of thought~\cite{w3c_data_integrity_2025}, and (iii) record its actions as signed, tamper‐evident logs bound to Decentralized Identifiers~\cite{w3c_did_core_2022}.  These capabilities provide cryptographic guarantees across the entire AI workflow, mitigating hallucinations and safeguarding against spoofed or malicious inputs.

Emerging frameworks such as LangChain’s \texttt{id\_transformer} and Microsoft’s Semantic Kernel already offer built‐in hooks for performing these verifications and generating proofs during prompt dispatch and output handling, indicating that the Semantic Web’s Proof \& Trust layers may finally see practical uptake in next‐generation autonomous AI systems.

\subsection{Lessons from the Forgotten AI Wave}

The Semantic Web’s history offers valuable lessons on AI’s cyclical evolution. The shift from logic-based AI to deep learning was not a rejection of earlier methods but rather an adaptation to technological constraints and industry priorities. As AI progresses, integrating structured knowledge and its cryptographic proofs with modern techniques could yield more powerful, interpretable, and autonomous systems.

Recognizing the Semantic Web as part of AI’s history provides a more comprehensive view of technological evolution. Instead of seeing it as a failed wave, we can understand it as a crucial step in AI’s development—one whose principles may yet play a key role in the next generation of intelligent systems.

\section{Conclusion: Reintegrating the Semantic Web into AI's History} \label{sec:conclusions}

Our key contributions in this paper are as follows:

\begin{itemize}
    \item We reviewed previous illustrations of AI history and synthesized them into a uniform scale in Section~\ref{sec:ai-history}.
    
    \item We identified the Semantic Web and Software Agents as a forgotten wave of Artificial Intelligence by connecting them to the current AI boom and analyzing academic statistics, as discussed in Section~\ref{sec:introduction} and Section~\ref{sec:academic-statistics}.
\end{itemize}

The Semantic Web represents a significant yet often underappreciated chapter in the evolution of artificial intelligence. Emerging in the early 2000s, its vision of a machine-readable web, capable of empowering AI-driven Software Agents, promised transformative applications ranging from intelligent automation to advanced data integration. While the grand vision articulated by Tim Berners-Lee, James Hendler, and Ora Lassila remains unrealized in its entirety, its influence on structuring and interpreting web data continues to resonate in modern technologies, including knowledge graphs, ontologies, and interoperability standards.

In this paper, we revisited the academic and industrial impact of the Semantic Web, using bibliometric analysis and historical context to argue for its rightful place within AI’s broader narrative of hype cycles, AI winters, and eventual breakthroughs. In particular, the overlooked connection between the Semantic Web and AI Software Agent research from 2000 to 2010 highlights its contributions to advancing intelligent systems. Despite challenges in achieving widespread adoption and the subsequent shift in focus toward machine learning and deep learning paradigms, the Semantic Web established foundational principles that remain relevant to contemporary AI development.

Recognizing the contributions of the Semantic Web enriches our understanding of AI’s history and underscores the importance of examining past technological waves to draw lessons for the future. By reintegrating this forgotten wave into AI’s historical discourse, we gain a more comprehensive perspective on the cyclical nature of innovation and the interplay between vision and practicality in shaping technological progress.


\newpage

\bibliographystyle{apalike}
{\small
\bibliography{references_short}}

\end{document}